\documentclass[12pt,preprint]{aastex}

\begin{document}

\title{L-dwarf variability: Magnetic star spots or non-uniform clouds?}

\author{Christopher R. Gelino, Mark S. Marley\altaffilmark{1}, Jon A.
Holtzman}
\affil{Department of Astronomy, New Mexico State University, Las Cruces,
NM, 88003}
\email{crom@nmsu.edu, mmarley@mail.arc.nasa.gov, holtz@nmsu.edu}

\author{Andrew S. Ackerman}
\affil{NASA Ames Research Center, Moffett Field, CA, 94035}

\and

\author{Katharina Lodders}
\affil{Planetary Chemistry Laboratory, Department of Earth and Planetary
Sciences, Washington University, St. Louis, MO, 63130}

\altaffiltext{1}{NASA Ames Research Center, Moffett Field, CA, 94035}

\begin{abstract}
The recent discovery of photometric variations in L dwarfs has
opened a discussion on the cause of the variations.  We argue against the
existence of magnetic spots in these atmospheres and favor the idea that
non-uniform condensate coverage (i.e. clouds) is responsible for the
variations.  The magnetic Reynolds
number ($R_m$) in the atmosphere of L dwarfs, which describes how well
the  gas couples with the magnetic field,  is too small ($\ll$1) 
to support the formation of magnetic
spots.  In constrast silicate and iron clouds form in the photospheres
of L dwarfs. Inhomogeneities in such cloud decks can plausibly
produce the observed photometric variations.  
Further evidence in support of clouds is the tendency for
variable L dwarfs to be bluer in $J-K_s$ than the average L
dwarf of a given spectral type.  This color effect is expected if clear
holes appear in an otherwise uniform cloud layer.
\end{abstract}


\section{Introduction}
Like stars of earlier spectral types, some L dwarfs \citep[the spectral
type cooler than M;][]{kir99,mart99,bas00,kir00} are variable
\citep{tin99,bai99,bai01}.   In this Letter we examine whether magnetic
`star' spots or clouds are more likely to be responsible for the
observed variability.  We conclude that patchy clouds are the most
plausible mechanism for the observed variability and discuss how they
can produce the three types of variations seen: 1) periodic, 
2) periodic but periods change on long time scales (months), and 
3) non-periodic.

\citet{tin99} presented the first attempt to detect clouds in brown
dwarf atmospheres.  They observed the M9 brown dwarf LP944-20 and the L5
brown dwarf DENIS 1228-15 through two narrow-band filters designed
to detect changes in TiO absorption.  The changes in the TiO band
strength were presumed to indicate changes in the opacity, which occurs
when TiO is depleted through condensation.  They found that
LP944-20 was variable, but DENIS 1228-15 was not.  The
authors speculated that the passage of clouds over the disk of LP944-20
produced small changes in the effective temperature that caused the
small
variations ($\sim$0.04 magnitude) in their narrow-band filters.  These
results do not exclude the possibility of clouds in their L dwarf,
however,
since variations of this amplitude would have been difficult to detect
given that the errors for that object were larger than 0.04 magnitude.

\citet{bai99} conducted a variability search 
in the broad-band $I$ filter and found evidence of variability in
the L1.5 dwarf 2MASS 1145+23.  The object displayed $\sim$0.04
magnitude variations that repeated with a period of 7.1 hours.
In an expanded study of 21 L and M dwarfs
\citet{bai01} found that
over half of their sample exhibited statistically significant variations
with amplitudes 0.01 to 0.055 magnitude and time scales of 0.4 to 100
hours.  They were unable to find periodic light curves for many of the
variables. 2MASS 1145+23, however, now exhibited
variability with a period of 11.2~hrs.
A similar varying period has also been observed in an M9.5 dwarf
star by \citet*{mart01}, who suggested that
evolving surface features, possibly dust clouds or
magnetic spots, were responsible for the change.  

At first glance, either magnetic spots or clouds could plausibly
be associated with the observed variability.  Magnetic fields have
been measured for several M dwarfs and estimates of their strength
are a few kG \citep{saar94,joh96}.  The surface filling factors 
are generally $>50\%$.  Unfortunately, periodic photometric
variations have been observed in only
a few M dwarfs, suggesting that either the surfaces of these objects are
completely covered with spots or that the spots are few in number and
uniformly distributed \citep*{haw00}.  In addition, \citet{bon95}
reports that no good data exist to support the scenario of cyclic,
organized spots in objects cooler than M0.  

For the L dwarfs, clouds are also a reasonable potential source
of variability.  Iron, enstatite, and forsterite are
the most abundant species expected to condense at the atmospheric
temperatures and pressures characteristic of the L dwarfs 
\citep{lod99,bur99}.
Once condensed, the species likely settle into discrete, optically
thick, cloud decks, with optically thicker clouds arising in progressively
later L dwarfs \citep{marl00,ack01}.  Since
the atmospheric circulation pattern of most
L dwarfs is likely similar to that of Jupiter
\citep{sch00} it is not unreasonable to expect that many
L dwarfs also have a banded appearance.  Any large  inhomogeneities
(thicker clouds or holes in the cloud deck) could then produce
a photometric signal.  \citet{gel00} have shown that if Jupiter was an
unresolved point source, the Great Red Spot would provide a
photometrically detectable signal.  If L dwarfs have similar cloud
features to Jupiter, then it is plausible that they may also
exhibit photometric variations.  The observation that more of the 
later L dwarfs are variable than the earlier type \citep{bai01}
 supports the scenario of cloud-caused variations.

In addition to the photometric variability, some L dwarfs exhibit
H$\alpha$ emission \citep{kir99,kir00}, which is known to be an 
indicator of high chromospheric temperatures and magnetic 
activity in earlier type stars \citep{haw00}.  If the variability
in L dwarfs is caused by magnetic spots, then it is plausible to
expect a correlation between H$\alpha$ emission and the variable
objects.  Gelino et al. (in preparation) have combined their 
L-dwarf photometric
variability results with those of \citet{bai01} and find no 
correlation between H$\alpha$ emission (i.e. magnetic activity) and
variability, supporting the  conclusions of \citet{bai01} and
\citet{mart01}.  

As with earlier-type stars, wave heating has been suggested as
the mechanism responsible for producing hot brown dwarf upper
atmospheres \citep{yel00}. Convection-caused waves propagate
and grow as they rise through the upper atmosphere, eventually 
releasing their energy and heating the gas as they dissipate. 
These high temperatures
combined with magnetic field effects are likely responsible for
the  H$\alpha$ emission.  L dwarfs with weak magnetic fields should
exhibit little or no H$\alpha$ emission.  In light of this it is
quite notable that the fraction of objects with H$\alpha$ emission
peaks at spectral type M7 and decreases at earlier and later spectral
types \citep{giz00}.  No L dwarfs later than L5 show
H$\alpha$ in emission \citep{kir01}, although one T dwarf does 
\citep{burg00}.  Since the early L population consists of both
young brown dwarfs and old stars, this trend could indicate an inability
of either substellar objects or cool stars to produce
magnetic fields appropriate to maintain the H$\alpha$ 
emission \citep{giz00}.

Using kinematics as a probe for age, \citet{giz00} have argued that the
old stellar L-dwarf population is more likely to show H$\alpha$
emission than the younger, lower mass population.  This might imply that
the process by which H$\alpha$ emission is produced is driven by the
mass of the object and not the effective temperature.  Indeed, the
dissipation of acoustic waves in the upper atmosphere could be
responsible for the heating of the chromosphere.  Unfortunately, this
process is poorly characterized at the masses and effective temperatures
of interest here.

In addition to H$\alpha$ emission, radio emission can also be a
signature of a magnetic field.  \citet{ber01} recently reported the
detection of a radio flare as well as quiescent emission from 
the M-dwarf LP944-20.
They infer that the radio emission is caused by synchrotron emission
and estimate a field strength of $\sim$5 G, much less than the
field strengths of active M dwarfs \citep{hai91}.  In addition the
substellar
nature, old age \citep{tin98a}, and rapid rotation \citep{tin98b} all
support the weak field strength; many L dwarfs with
spectroscopically determined rotation velocities are rotating quite
rapidly and lack significant H$\alpha$ emission \citep{haw00},
suggesting that the magnetic fields of these presumably old objects are
too weak to slow down the rotation.

The existence of a magnetic field in LP944-20 is also supported by the
observation of an X-ray flare \citep{rut00}.  An X-ray flare in an old,
non-accreting object such as this can only be caused by magnetic
activity.  However, the lack of quiescent X-ray emission suggests
that the magnetic field is quite weak.  Rutledge et al. postulate
that because of the rapid rotation in this object, either 
the turbulent dynamo
is being suppressed or the magnetic field is being configured into a
more organized form.  They also suggest that the lack of ionization in
the cool photosphere prevents the magnetic field from coupling with the
gas in the atmosphere, causing the field to dissipate.

\citet*{fle00} arrive at the same conclusion with their study of the
X-ray flare of the M8 dwarf VB 10.  They go so far as to estimate the
ionization fraction in its atmosphere, by extrapolating the ionization
fractions in the atmospheres of early dwarfs down to the effective
temperature of VB 10.  They estimate that the ionization fraction in
late M dwarfs is 2 orders of magnitude lower than in early M
dwarfs and 3 orders of magnitude smaller than in the sun, and conclude
that that magnetic footprints (i.e. spots) are unlikely to exist in 
the photosphere.  By way of comparison, the cooler M9 dwarf LP944-20
is even less likely to have magnetic spots than VB 10, supporting
 the conclusion by \citet{tin99} that they were
detecting the signature of clouds.  This, in turn, implies that
magnetically produced spots are unlikely to be formed in L dwarfs and
the variations seen by \cite{bai01} and Gelino et al. (in preparation)
are caused by clouds.

In this Letter we further explore the possibility that the magnetic
field and atmospheric gas in L dwarfs are not coupled.  We improve 
on the estimations of \citet{fle00} by utilizing models of the 
abundances of ions and neutrals appropriate for L-dwarf atmospheres.

\section{Magnetic Spots}
Magnetic spots in the sun (i.e. sunspots) are thought to form
as magnetic flux tubes rise to the photosphere \citep{par55}.  For
magnetic buoyancy to be important, the plasma must be a sufficiently
good conductor.  This criterion is most likely not satisfied in
cool L-dwarf atmospheres, especially in the low pressure regions where
the temperatures are also low.  In these regions the free electron
abundance is extremely small ($<0.1$ cm$^{-3}$ above 1 bar 
for a 2000 K model), therefore, the regions do
not make good conductors.  Consequently, it would seem that this
condition alone prevents the formation of any magnetic spots.  

To estimate the strength of coupling between the gas and the magnetic
field we compute the magnetic Reynolds number $R_m$ for L-dwarf 
atmospheres.  $R_m$ is a dimensionless parameter
describing how efficiently a gas interacts with a magnetic field;
$R_m=lv/\eta$ \citep{pri82}, where $l$ is a length scale, $v$ is
a velocity scale,
and $\eta$ is the magnetic diffusivity of the gas.  When $R_m\ll$~1, 
the magnetic field slips through the gas with no interaction; for
$R_m\gg$~1, the magnetic field is frozen in the gas.

To compute $R_m$ we rely on atmosphere models computed by
\citet{marl01} for cloudy L dwarfs.  The models employ the
cloud model of \citet{ack01} with the cloud parameter
$f_{\rm rain} = 5$.  The L-dwarf $T_{\rm eff}$ range is still uncertain,
but likely lies between about 2200 and 1300 K.  We consider models with
$T_{\rm eff}$ of 2000 to 1200 K and surface
gravity of $10^5$ cm s$^{-2}$, appropriate for a $\sim 35$ Jupiter
mass brown dwarf.

The appropriate length scale $l$ to use in the calculation of
$R_m$ is not obvious.
In the area around sunspots, $l$ is usually
taken as the size of the sunspot, a small fraction of the
solar radius.  We choose to set $l=H$, the pressure scale height.
Typical values of $H$ are around 10$^6$ cm
at the 1 bar level.
For the velocity scale we use the convective velocity of the atmosphere,
$w_*$, as computed in \citet{ack01} and references therein.  Ackerman
\& Marley limit the size of $w_*$ above the radiative-convective
boundary.  To compute an upper limit to $R_m$ we compute
$w_*$ assuming that the entire thermal flux of the brown dwarf
is carried by convection throughout the atmosphere.
Typical convective velocities are computed to be 
10$^3$-10$^4$ cm s$^{-1}$.

The value for $\eta$ is computed from \citet{pri82};
$\eta = 5.2 \times 10^9~\ln \Lambda~T^{-3/2}~A~{\rm cm^2\,s^{-1}}$,
where
$A \approx 1 + 5.2 \times 10^{-11} (n_n/n_e) (T^2/\ln \Lambda)$
is a factor to account for the partial ionization of the plasma, $T$
is temperature, $n_n$ is the number density of neutral
atoms and molecules, $n_e$ is the number density of electrons, and $\ln
\Lambda$ is the Coulomb logarithm.
We use the abundance tables of \citet{lod99} to calculate $n_e$ and
$n_n$.  The appropriate value of the Coulomb
logarithm, however, is not known for an L-dwarf atmosphere.  Table 2.1
from
\citet{pri82} gives the value of $\ln \Lambda$ for temperatures and
densities relevant for the sun.  Unfortunately, this table provides
no entries at the densities and temperatures needed for the
extrapolation
to the parameters of an L-dwarf atmosphere since $\ln \Lambda$ becomes
small
($\lesssim 5$).  However, \citet{chen74} comments that it is sufficient
to
use $\ln \Lambda$=10 regardless of the specific plasma parameters.  This
is
well supported from the data in the table in Priest since $\ln \Lambda$
only varies from $\sim5-25$ over 15 orders of magnitude in density and 3
orders of magnitude in temperature.  We use a value of 5 for our
calculations, but test the sensitivity of $R_m$ on our somewhat
arbitrarily chosen value by computing $R_m$ with $\ln \Lambda$ at the
unrealistic values of 1 and 100.
We find that $R_m$ computed at these limits differs by $<10^{-4}\%$ for
our coolest model and $<1\%$ for our warmest model and stress
that for the region of parameter space covered by our calculation
the precise value of $\ln \Lambda$ is essentially irrelevant.

The value of $R_m$ as a function of pressure in the model atmospheres is
shown in Figure 1.  $R_m$ is very small throughout the entire upper
atmosphere; only at pressures of $\sim 100\,\rm bar$ and higher does
$R_m$ start to approach 1.  By comparison, $R_m$ near sunspots is
estimated to
be 10$^4$-10$^6$ \citep{pri82}.  Open circles denote the approximate
base of the photosphere (where $T=T_{\rm eff}$) for the models shown.  Note 
that in contrast to the sun, $R_m$ only approaches unity well below the
photosphere.
At $R_m$=1, the plasma and the magnetic
field will interact with each other only to a small degree.  
Any weak magnetic disturbances deep in these atmosphere
are unlikely to affect the
surface thermal flux since the winds and weather patterns alluded
to earlier will redistribute the upwelling thermal flux before
it is radiated.   Thus, throughout the atmospheres of L dwarfs,
spanning the range from roughly L2 to L8 \citep{kir99,kir00} we expect
little or no interaction between the atmosphere and the magnetic field.

\section{Clouds}
The thermal fluxes emerging from L-dwarf atmospheres are
affected by clouds.
Early L dwarfs have relatively thin clouds high in the
atmosphere \citep{ack01}.  Figure 1 illustrates the trend for 
the clouds to form progressively
deeper in the atmosphere at later spectral types.  Again the
base of the photosphere is marked for the models shown.
Unlike the case
for $R_m$, the clouds form in the immediate vicinity of the photosphere
and thus are well placed to affect the emitted thermal flux. 
For objects with
$T_{\rm eff} \le2000 K$ \citep[approximately L2 and cooler;][]{ste01}
clouds play an important role in the emitted flux since the 
condensate opacity is significant.

For an arbitrary L dwarf the emitted flux in some spectral regions
will be limited by the cloud deck, while in others gaseous opacity
reaches optical depth unity above the cloud.  \citet{marl01}
illustrate the effects of gaseous and condensate opacity for
a variety of L dwarfs.  If there is a transitory clearing in the cloud
deck additional flux will emerge from those spectral regions in which
the cloud opacity is otherwise dominant.  Examples are the peaks of flux
emerging from the water band windows in {\it z, J, H,} and $K$ bands and
the optical flux in $I$ band.  The resulting bright spots on the
objects will be similar to the `5-$\mu$m hot spots' of Jupiter
\citep{wes74} where flux emerges from holes in the ammonia cloud.
An atmosphere with such non-uniformly distributed high contrast regions
is quite capable of producing photometric variations.

As cloud optical thickness increases with later spectral type, 
L-dwarf $J-K_s$ color becomes redder,
eventually saturating around 2 \citep{marl00,all01,ack01,marl01,tsu01}.
Clouds in cooler objects lie well below the photosphere, and 
leave the radiating region in the atmosphere 
relatively clear.  The clear atmosphere partially manifests itself
in the blue $J-K_s$ seen in the T dwarfs \citep{all96,marl96,tsu96}.

These trends suggest that infrared colors of variable objects may
support the cloud model for variability.  
The $J-K_s$ color of those objects which have been surveyed for
variability is shown in Figure 2.  Although subject
to small number statistics, suspected variables with spectral
types later than L2 tend to be bluer than the average L dwarf \citep{kir00}
at the same spectral type.  Models predict \citep{marl00,marl01} 
that a hypothetical L dwarf with no clouds will be substantially bluer
\citep[$\sim1.5$ mag;][]{marl01b}
at $J-K_s$ than a more realistic object with the same effective
temperature and a cloudy atmosphere.  Thus, if the average L dwarf at
a given spectral type is entirely covered with clouds, it would
not likely be seen as a variable and it would have more typical 
$J-K_s$ color.  In order for photometric variations to arise by
the cloud mechanism there must be
non-uniformity in the cloud coverage, such as clearings in the clouds.
So, not only would clear sections of the atmosphere (holes) provide a source
for brightness variations, flux emerging through the hypothetical
holes would cause the $J-K_s$ color to be somewhat bluer
than the average object.  The lack of a similar trend for the
early L and late M dwarfs might indicate a different mechanism is
at work in those atmospheres.  Clearly more data, including time
resolved multi-color photometry, are required to determine if variability
is indeed connected to color.

\citet{sch00} show that the atmosphere of L dwarfs can exist in one of two
states, chaotic or banded.  Since clouds respond to atmospheric motions,
they would presumably reflect one of these two morphologies.
 In general, higher mass objects will have more
chaotic and three-dimensional internal dynamics than lower mass objects,
meaning that the higher mass objects are less likely to have banded
cloud features.  It is not obvious which cloud morphology would better
produce photometric variations.  For example, objects with more
chaotic atmospheres might be more likely to have uniformly distributed
clouds. When rotating such objects might show little photometric variation.  
If the chaotic atmospheres result in fairly
complete cloud coverage, then we might expect that more massive L
dwarfs would be redder in $J-K_s$ and tend not to be variable.  
Conversely, if chaotic atmospheres more often produce large
holes in the clouds they may more easily produce photometric
signatures than banded atmospheres.  In this scenario, the more
massive L dwarfs would be bluer in $J-K_s$ and tend to be variable.
Furthermore, rapidly evolving, chaotic
atmospheres may be responsible for the changes in photometric
period observed for some objects.  It is easy to imagine similar
scenarios for banded clouds.  

The presence of clouds can also explain the different types of variability
seen in L dwarfs.  For example, periodic variations over time scales 
of days can be explained by a long-lived clearing or thickening in the
clouds.  Rotation modulation of the photometric signal will then produce
a periodic signal.  The inhomogeneities could migrate 
latitudinally or dissipate and reform at a different latitude
as does the Great Dark Spot in the atmosphere of Neptune \citep{ham97}.  
If wind speed is a function of latitude as on all the giant
planets of our solar system, then spots
at different latitudes will circle the object with different periods.
Observations taken several months apart
will show periodic variations with different periods as is reported 
for the L-dwarf 2MASS 1145+23.  Finally, L dwarfs whose clouds 
change on time scales of a few days or less or whose cloudy spots
are distributed at several latitudes with different wind speeds could still produce
a 
photometric signal, but the signal might not be periodic.
Clearly, more observations and modeling are required
to better characterize atmospheric circulation and weather in
L-dwarf atmospheres.

\section{Conclusions}
There is little doubt that some L dwarfs are photometrically variable.
Although most authors suggest that the variations are
caused by clouds, the possibility of magnetic spots is often
mentioned \citep[e.g.][]{bai99,bai01,mart01}.  We have shown that the low
ionization fraction predicted by L-dwarf models and the accompanyingly
low magnetic Reynolds numbers strongly argue against
spots as a possible cause for the photometric variations.  
On the other hand silicate and iron grains condense in L-dwarf atmospheres 
within the photosphere.  These clouds are likely responsible for the 
photometric variations discovered in the various studies, particularly for
the later L dwarfs (about L2 and later).  Since the thermal emission
of T dwarfs is also influenced by clouds (Marley et al. 2001) we predict
that variability will also be found in the opacity window regions of these
objects.
Further work with models and more observations are clearly needed to better
understand cloud composition and dynamics.

\begin{acknowledgements}
The authors wish to thank Jason Peterson for his help with Figure 1.
C.G. and M.M. acknowledge support from NASA grants NAG2-6007 and NAG5-8919
and  NSF grants AST-9624878 and AST-0086288. 
Work by K.L. supported by NSF grant AST-0086487.
\end{acknowledgements}

\begin{figure}
\plotone{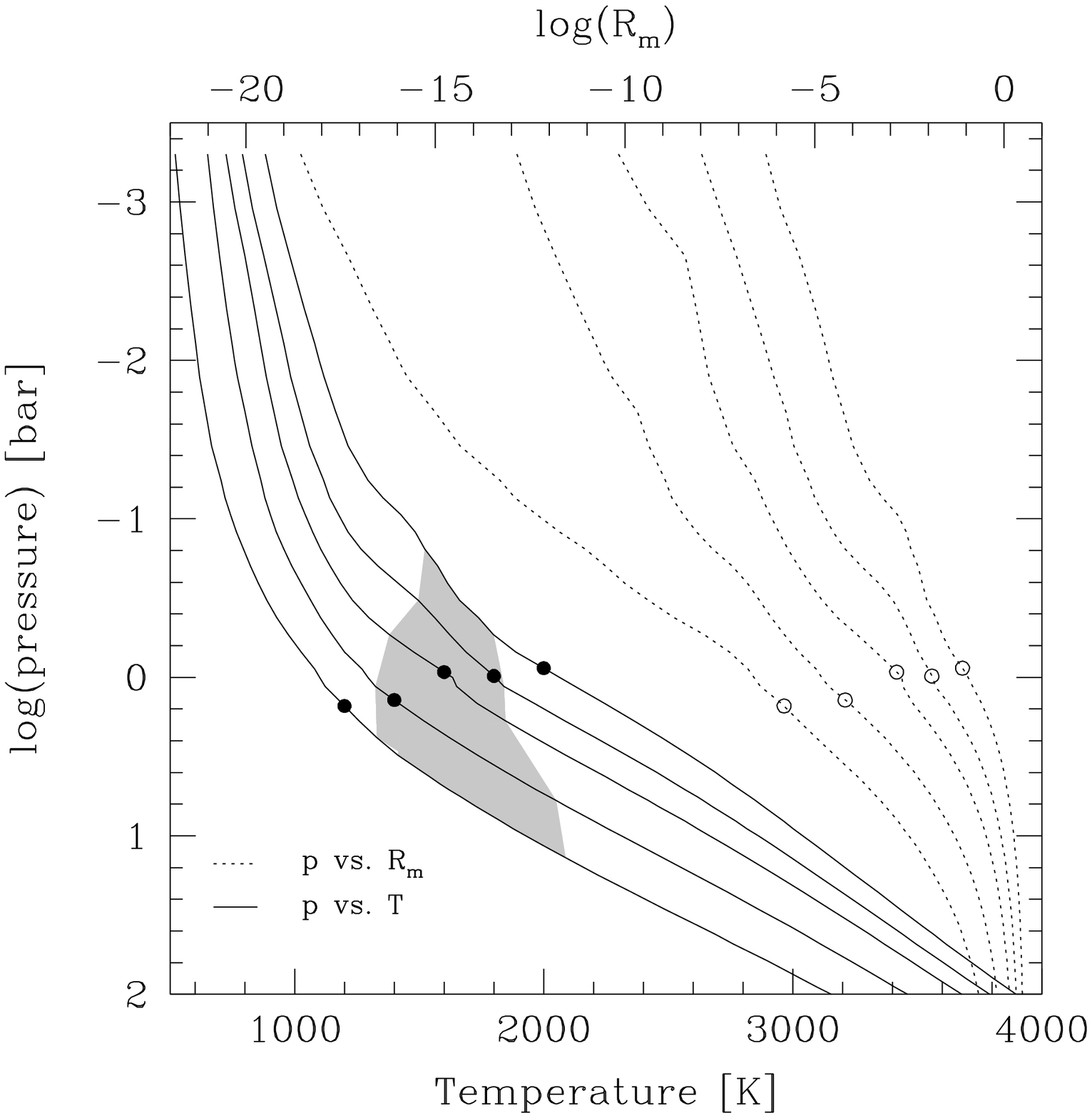}
\caption{Magnetic Reynolds number ($R_m$, dotted lines) and temperature
(solid lines) plotted as a function of pressure in the atmospheres of
L-dwarfs with $T_{\rm eff} = 1200$, 1400, 1600, 1800, \& 2000 K,
going from left to right.  The solid dots represent the pressure at
which
the temperature matches $T_{\rm eff}$ (approximately the photosphere);
open circles are the photosphere location in p vs. $R_m$ space;
the shaded area is the
region from the cloud bottom to cloud top (defined here as the level where
the cloud's optical depth is $\sim0.1$).  $R_m$ is quite small throughout 
the entire atmosphere
and starts approaching 1 below the base of the clouds 
and the level of the photosphere. }
\end{figure}

\begin{figure}
\plotone{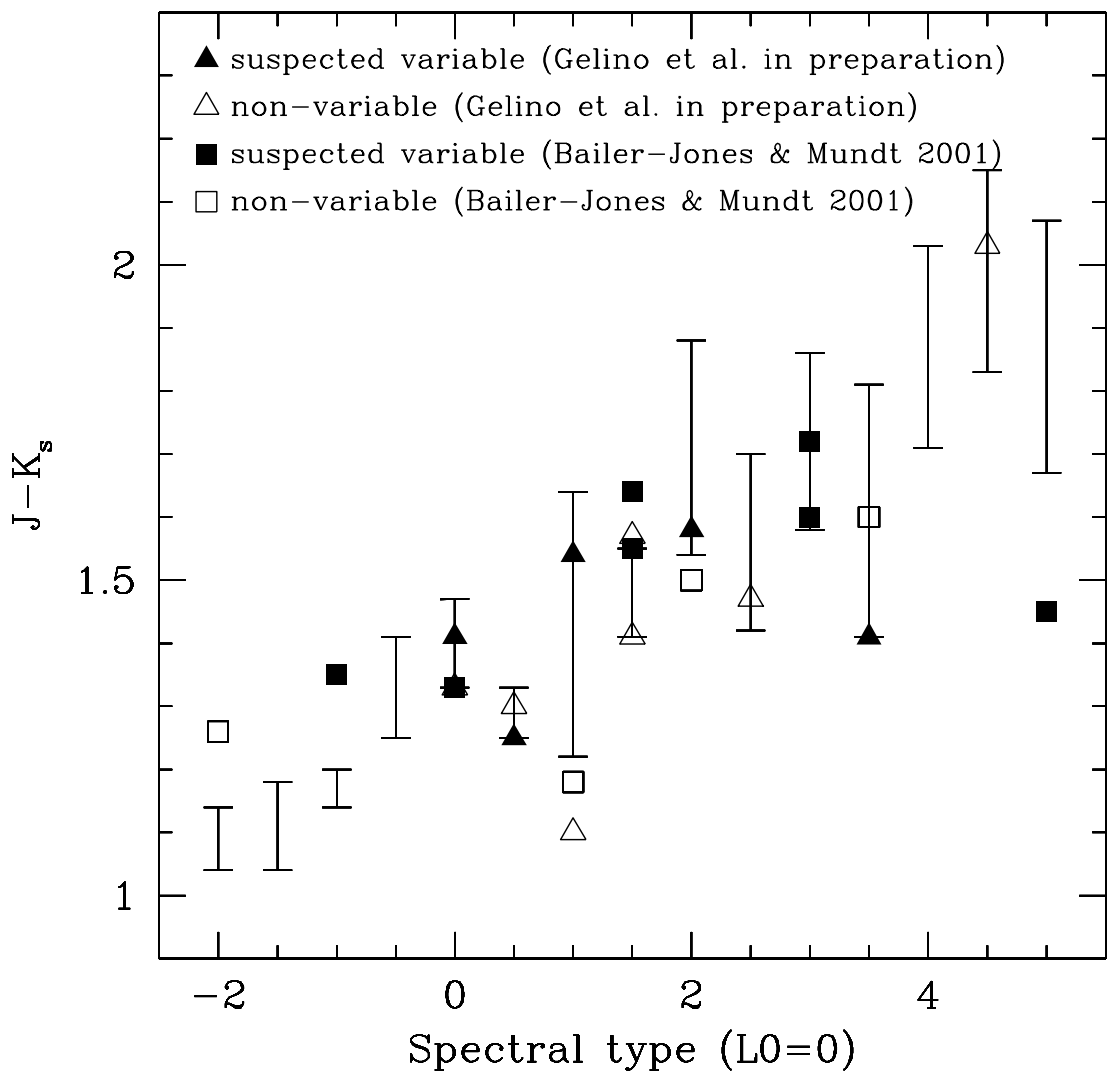}
\caption{$J-K_s$ as a function of spectral type for the \citet{bai01}
and Gelino et al. variability searches.  Also shown are the
$\pm$1-$\sigma$ spread of $J-K_s$ colors (error
bars) for spectral
types M8-L5 \citep{kir00}.  The
suspected variables (solid points) with spectral types L2 and later are all
at or bluer than the average $J-K_s$ for that spectral type.}
\end{figure}

\end{document}